\documentclass[pra,aps,longbibliography,showpacs]{revtex4-1} 
\usepackage{silence}
\usepackage{amsmath,revsymb,graphicx,amssymb,array,natbib,multirow,adjustbox}

\newcommand{\kt}[1]{\ensuremath{|#1\rangle}}

\newcommand{\HS}{\mathcal{H}}

\begin{document}

\title{Symmetry, Structure, and Emergent Subsystems}

\author{N.L.~Harshman\email{Electronic address: harshman@american.edu}}

\affiliation{Department of Physics, 
4400 Massachusetts Ave.\ NW, American University, Washington, DC 20016-8058}

\begin{abstract}
Symmetries impose structure on the Hilbert space of a quantum mechanical model. The mathematical units of this structure are the irreducible representations of symmetry groups and I consider how they function as conceptual units of interpretation. For models with symmetry, the properties of irreducible representations constrain the possibilities of Hilbert space arithmetic, i.e.\ how a Hilbert space can be decomposed into sums of subspaces and factored into products of subspaces. Partitioning the Hilbert space is equivalent to parsing the system into subsystems, and these emergent subsystems provide insight into the kinematics, dynamics, and informatics of a quantum model.  This article provides examples of how complex models can be built up from irreducible representations that  correspond to `natural' ontological units like spins and particles. It also gives examples of the reverse process in which complex models are partitioned into subsystems that are selected by the representations of the symmetries and require no underlying ontological commitments. These techniques are applied to a few-body model in one-dimension with a Hamiltonian depending on an interaction strength parameter. As this parameter is tuned, the Hamiltonian runs dynamical spectrum from integrable to chaotic, and the subsystems relevant for analyzing and interpreting the dynamics shift accordingly.
\end{abstract}

\maketitle

\section{Introduction}

One of the most evocative results in the whole history of mathematical physics that there are exactly five polyhedra with perfect symmetry, i.e.\ all faces, edges and vertices are congruent.  Starting from the intuitive and reasonable definitions and axioms of Euclidean geometry and applying the constraint of symmetry, these five structures are logically inevitable. Book 13, the climax of Euclid's \emph{Elements},  constructs these solids and proves they exhaust the possibilities. While the beauty of symmetric geometry provokes attention, I believe it is as much their `fiveness' that inspires. Plato, Kepler and others sought five-fold explanations for the physical structures of the world~\cite{wilczek_beautiful_2015}. Those chimerical hopes notwithstanding, Platonic solids do appear throughout science and technology for natural and practical reasons: they are fundamental structures.

Starting in the 19th century, combining generalized notions of geometry and symmetry yielded a bounty of new structures. In turn, these structures provided a framework for interpreting, classifying, and generating mathematical models of physical reality. Glossing over the technical details (upon which hang critical distinctions and academic careers), the same structural system that classifies Platonic solids and their generalizations in any dimension also classifies Lie groups, dynamical catastrophes, symmetric manifolds, random matrices, topological insulators, gauge quantum field theories, and so on. The imposition of symmetry on vectors spaces is surprisingly rigid, often giving finite or countable possibilities. And then, more often then not, it seems we find these possibilities manifest in our models of systems and dynamics. This coincidence of logically inevitable mathematical structures with elements of physical reality remains as seductive now as it did to Plato and Kepler. Some theoretical physicists spend their career chasing these beautifully symmetric models and enumerating their qualities, hoping one day the model will play an important role in the next `paradigm shift'.

My proposal is more modest: symmetry is not the answer to every question and the universe may have contingent features that no model can predict or explain. However, symmetries seem to exist in reality and certainly exist in many effective and productive models of reality. Within a framework like quantum mechanics, the presence of symmetry entails the existence of mathematical structures that are privileged by their relation to the symmetry. The focus of this article is the particular structures called irreducible representations of symmetry groups. Similar to how a Hamiltonian gives a spectrum in quantum mechanics, the irreducible representations (irreps) of a group of symmetry transformations form a kind of basis for possible manifestations of a symmetry on vector spaces. Like the Platonic solids, these irreps are the denumerable `atoms' of the Hilbert spaces of quantum mechanical models with symmetry.

For example, any quantum model with $\mathrm{SU}(2)$ symmetry has a description in terms of units that are often called spins, even when they have no rotational origin. These spin units are Hilbert subspaces that carry irreps of $\mathrm{SU}(2)$, and the total Hilbert space is reducible into products and sums of these `atoms' of spin.

Similarly, quantum field theory is built by associating free particles to irreps of the Poincar\'e group of symmetry transformations of Minkowski space-time, see for example \cite{streater_pct_1964, haag_local_1992, weinberg_quantum_1995}.  As Eugene Wigner discovered~\cite{wigner_unitary_1939}, the irreps of the Poincar\'e group are conveniently labeled by three invariants: mass, internal energy and spin~\footnote{Technically, relativistic particles correspond irreps of the universal covering group of the Poincar\'e group. This is necessary to account for the fact that in quantum mechanics we are typically interested in projective representations, i.e.\ representations up to a phase. Projective representations are more general than representations and they preserve the interpretation of pure states as rays in the Hilbert space. This technicality is mostly swept under the rug for the rest of this article.}. Irreps of the Poincar\'e group are equivalent to relativistic particles for most technical and interpretational purposes. A similar construction with the Galilean group and the non-relativistic particles is discussed below.

These examples make it clear that using irreps to analyze symmetry in quantum mechanics is an old story, full of many bold successes and productive technical details. Why rehash it here? The purpose of this contributed chapter is to explore the connections between the \emph{mathematical} units of symmetry embodied by irreps, arguably the `most inevitable' symmetric structures of quantum mechanics, and the \emph{conceptual} units of reality that form the basis for interpretation of quantum theories. Since irreps are symmetric structures that have the appealing properties of being denumerable, they hold the same appeal as Platonic `fiveness'. Their logical inevitability (given the standard formulation of quantum mechanics) make them the natural vocabulary for asking and answering questions about the fundamental nature of quantum reality, whether a more epistemic or ontic interpretation is advanced. At the very least, any interpretation of quantum mechanics must provide some justification for the `unreasonable effectiveness' of these mathematical structures as conceptual units. At the same time, the conceptual overreach perpetrated by natural philosophers enamored by the Platonic solids should serve as a cautionary tale.

With this motivation, and Sect.~II gives the local definitions of quantum mechanical models and symmetries, Sect.~III and IV subject the reader to a technical presentation about `Hilbert space arithmetic'. The Hilbert space can be built `bottom-up' out of products and sum of irrep spaces, or reduced `top-down' into products and sums of subspaces that carry irreps. I argue that Hilbert space arithmetic, and identifying how symmetry manifests itself in that arithmetic, provides a lens for understanding quantum concepts like energy levels, entanglement, locality (and its generalization specificity), distinguishability, and emergence.

But the view through that lens is not yet clear, so I also offer to the skeptical reader an application of symmetrized Hilbert space arithmetic. In Sect.~V, I investigate whether the mathematical structures of irreps have interpretative power when considered as conceptual units of reality within a model of quantum few-body systems in one dimension. This choice is partially due to the charming coincidence that such systems can carry realizations of the same symmetries as the Platonic solids and their generalizations. More meaningfully, quantum few-body systems in one dimension are at the knife's edge in terms of dynamical regimes for closed quantum systems. Depending on the symmetry of the Hamiltonian, such systems can have dynamics that are regular, integrable and solvable or irregular, chaotic, and ill-conditioned. I argue that irreps form conceptual units to interpret this rich physics of few-body systems.

The stakes of this analysis have been raised by recent experiments with ultracold atoms in effectively one-dimensional optical traps~\cite{serwane_deterministic_2011, zurn_fermionization_2012, zurn_pairing_2013, wenz_few_2013}. These offer the possibility of implementing controllable few-body models and they provide a relevant milieu for evaluating claims of interpretative utility of irreps as conceptual units. Because irreps are derived from symmetries of the Hamiltonian of a few-body model, they can be used to find useful observables to describe the collective degrees of freedom of few-body systems~\cite{harshman_one-dimensional_2016,harshman_one-dimensional_2016-1}. Further, the connections between reducibility into irreps and solvability, while not completely clear, can still be productive in the quest to identify new solvable models in few-body physics. See \cite{harshman_infinite_2017,andersen_hybrid_2017} for recent examples. Finally, the connections between solvability and controllability provide opportunities for technological applications with ultracold atoms. Controllability is a key requirement for the relativity of entanglement pioneered by Zanardi (see Sect.~\ref{sec:zanardi} below). As a practical concern for developing quantum information processing devices, it motivates the search for emergent subsystems~\cite{zanardi_universal_2004}.

\section{Quantum Mechanical Models and Symmetry}

Symmetries are often invoked as grand, unifying principles at the foundations of quantum mechanics. For example, the Standard Model is presented as though it emerges as a logical consequence of combining the symmetries of Minkowski space-time with gauge symmetries of internal flavors. Such a presentation is an exaggeration, as the numerous unexplained free parameters in the Standard Model attest. Here I want to present symmetries in the narrower scope: symmetries as defined within the context of a specific quantum mechanical model.

\subsection{Model definition and scope}

Following Haag~\cite{haag_local_1992}, I define a quantum mechanical model as a set of observables $\mathcal{A}$ that are represented by Hermitian operators acting on a Hilbert space $\HS$. This rather airless definition of quantum mechanical model makes minimal ontological claims beyond the standard framework of quantum mechanics: pure states are unit vectors in $\HS$, measurements are expectation values of observables, Born rule and all that. The origin of the observables and their representation within the model could be established by scientific utility alone, such as making predictions, or by other epistemic concerns. Using this definition of model, I am consciously attempting to restrict my ontological commitment (as much as possible) to a structure of relatively-defined mathematical objects while still doing `normal' quantum mechanics. To those who have ontological commitments in their interpretations of quantum mechanics, I believe that this framework for models is general enough to encompass those theories, and so I ask indulgence in this exercise in structuralism.

The simplest non-trivial example of a model is given by the algebra of Pauli matrices acting on $\HS\sim\mathbb{C}^2$. Any two-level system can be described in this framework, and although there is an analogy to a spin-$1/2$ system, the model makes no ontological commitment to any fundamental particle or other unit of reality. The model could describe an isolated discrete degree of freedom, like the isospin of a nucleon, or could emerge as an effective theory, like the lowest two energy levels of a more complicated system. It could also be derived from the smallest non-trivial projective representation of an underlying three-dimensional rotation group, i.e.\ a true spin. In any case, the model as a mathematical structure can be studied without reference to any physical embodiment, ontological commitment, or larger theoretical framework.

Non-relativistic few-body models often do start with an ontological commitment, such as $N$ particles on an underlying $d$-dimensional Euclidean space $\mathcal{X}\sim\mathbb{R}^d$. The Hilbert space is realized as $\HS\sim L^2(\mathbb{R}^{Nd})$, Lebesgue square-integrable functions on the configuration space $\mathcal{X}^{\times N}$ constructed from the $N$-fold Cartesian product of single-particle spaces. The corresponding set of observables is generated by the Heisenberg algebra of $2Nd$ canonical position and momentum operators which are represented as variable multiplication and differentiation operators on $L^2(\mathbb{R}^{Nd})$.  There are certainly technical demons (like unbounded operators) hidden in this model, but for now let us assume they can be exorcised.

The subsequent analysis presumes the existence of a model, whether it has been defined in the abstract or derived from underlying particles or other units of reality, . The goal is then to use symmetry to identify structures within the model that aid interpretation of empirical phenomena without making further commitments. Note that in both of the examples presented above, the set of observables was actually an algebra of observables, closed under addition and multiplication. That will not always be the case for models I consider. For models with finite-dimensional Hilbert spaces the distinction is minor, but it is important for infinite-dimensional spaces.

In order to make progress, I restrict my attention to models with two additional properties. First, I assume that within the set of observables there is a particular observable $H\in\mathcal{A}$ called the Hamiltonian. The Hamiltonian generates time translations by the unitary operator $U(t) = \exp(-iHt/\hbar)$ defined on all of $\HS$. This assumption thereby excludes models that describe systems with time-dependent Hamiltonians or open dynamics. These are obviously cases of physical interest, but the complications they introduce require attention to detail beyond the scope of this article. 

Second, I assume that the set of observables is represented on a Hilbert space that is separable in the topological sense. A separable Hilbert space has a countable orthonormal basis~\cite{streater_pct_1964}. Equivalently, it implies  the existence of a decomposition of the Hilbert space $\HS$ of the system into a direct sum of a countable (but still possibly infinite) set of one-dimensional subspaces $\HS_i$:
\begin{equation}\label{eq:HSdecomp}
\HS = \bigoplus_i \HS_i.
\end{equation}
Restricting to models with separable Hilbert spaces is a convenience that (to my knowledge) is not much of a restriction for models of quantum mechanical systems with a finite number of particles. However, there are many-body systems and quantum field theories where this kind of separability cannot be assumed~\cite{streater_pct_1964}.

\subsection{Model symmetries}

Following Wigner~\cite{wigner_group_1959}, a symmetry of a model is a group of unitary (and possibly antiunitary) operators on $\HS$. Such operators preserve the magnitude of inner products and therefore of probabilities. For finite dimensional Hilbert spaces $\HS\sim\mathbb{C}^D$, the symmetry group of the model must be a subgroup of the unitary matrices $U(D)$. For infinite dimensional separable Hilbert spaces, the unitary group $U(\infty)$ can be defined by the inductive limit of $U(D)$~\cite{olshanski_problem_2003} and any subgroup of $U(\infty)$ could be a symmetry group. But most of these subgroups do not have any physical interpretation; they are purely formal. So how do we distinguish and classify useful or meaningful symmetry groups? 

Within the context of a formal model, the empirical content of a symmetry is inferred from its relation to the set of observables, specifically how the unitary group of symmetry operators transform the Hilbert-space representations of the observables. For example, the symmetry identifies invariant operators of the model, or more generally, can be used to classify operators based on their transformation properties~\cite{lombardi_role_2015}. A familiar example is classifying operators by their tensor rank under a representation of a matrix group, i.e. scalar, vector, pseudovector and so on for the rotation group.

However, starting from the formal model with Hilbert space and algebra and then identifying meaningful symmetries is usually not how physical analysis proceeds. The logic is usually reversed and the model is constructed from the symmetry of an underlying space or space-time. The group of unitary operators on the Hilbert space is a representation of the universal covering group of some space-time symmetry. The set of observables is the enveloping algebra built from the generators of the symmetry representation. Three relevant examples are: (1) spin models from rotational symmetry; (2) free non-relativistic particle models from Galilean symmetry~\cite{levy-leblond_nonrelativistic_1967}; and (3) free relativistic particle models from Poincar\'e symmetry~\cite{wigner_unitary_1939}. In each of these cases, the Hilbert space of a single spin or a single particle corresponds to an irreducible representation space of the space-time symmetry group~\footnote{As mentioned before, technically `particles' correspond to irreps of the universal covering groups of these space-time symmetries. This distinction allows for projective representations. Further, in order to get massive representations, an observable corresponding to mass must be added as a central element to the Galilean algebra.}. Bottom-up approaches to quantum mechanics often take these basic models as starting units, and build more complicated models out of their irreps spaces and algebras of observables.

In practice, identifying the symmetries of a non-relativistic few-body model takes place using a hybrid of formal, top-down and reductionist, bottom-up approaches. Symmetries in physical space are important, but so are symmetries in the spaces of configuration space and phase space. Although these auxiliary spaces can be derived from the bottom up approach starting with free particle models, the symmetries of these derived spaces may not be easily reducible into free-particle symmetries and their corresponding irreps can describe collective or emergent degrees of freedom. Additionally, few-body models can have symmetries that are defined by unitary operators on the Hilbert space itself without reference to a symmetry of an underlying space. For example, these kind of symmetries are present when there are accidental degeneracies~\cite{harshman_infinite_2017}.

\subsection{Kinematic and dynamic symmetries}

The symmetry groups with which I am primarily concerned are those represented by unitary operators that leave the Hamiltonian invariant. Call these \emph{kinematic symmetries} of the model. These unitary operators transform stationary states of the Hamiltonian, i.e.\ energy eigenstates, into other energy eigenstates with the same energy. Sometimes instead of considering a single Hamiltonian, it is profitable to consider the kinematic symmetry of a family of Hamiltonians that (for example) depends on a parameter or parameters but exist within the same model. This can be a tricky business, because sometimes as the parameter is varied the Hamiltonian loses the property of self-adjointness on the original Hilbert space of the model. A notable example is when parameter variation makes configuration space effectively disconnected~\cite{harshman_infinite_2017}. 

As opposed to the kinematic symmetry group of a Hamiltonian, a \emph{dynamic} symmetry group does not commute with the Hamiltonian as a whole (although some elements may). This definition encompasses a wide class of possibilities. In their simplest manifestation, dynamic symmetries provide algebraic relationships between the group and the Hamiltonian that map energy eigenstates into other energy eigenstates (like ladder operators) and induce algebraic relationships among expectation values of non-commuting observables~\cite{wybourne_classical_1974}. Poincar\'e and Galilean transformations are also dynamic symmetries in this sense; boosts change the energy of a state in an algebraic way.
Dynamic symmetries can also describe maps among Hamiltonians within a parametrized family, like a scale transformations~\cite{jackiw_introducing_1972}. They can even serve as maps between models and Hamiltonians that on the surface seem radically different, like supersymmetric partner Hamiltonians in quantum mechanics~\cite{cooper_supersymmetry_1995}.

\section{Hilbert Space Arithmetic: Direct Sum Decompositions}

The definition of model provided here seemingly relies on a the Hilbert space as an essential feature, and this section and the next are going to outline how Hilbert spaces can be decomposed into direct sums and factored into tensor products using observables and symmetries as the starting point. Topological separability of the Hilbert space is assumed throughout this vector space arithmetic, and the goal is to see how much interpretive power this kind of analysis can provide. Carving models into subspaces by decomposing into direct sum and into submodels by factoring into tensor products has a long history in quantum mechanics for practical reasons of mathematical analysis. I argue that it also forms units of conceptual analysis that are called different names in different contexts but are all manifestations of the same underlying structure. The tentative claim is that these structures are as valid as conceptual units of reality as are the more tangible concepts like `particles'. 

However, before proceeding, I must declare that I am not a Hilbert space fetishist. The specific topology of the Hilbert space is at once too loose and too restrictive for some purposes. For example, the domains on which the set of observables are bounded and self-adjoint may not be the entire Hilbert space, and their eigenspaces may not be contained in the Hilbert space. This technical aspect can be rigorously handled using Gel'fand triplets and rigged Hilbert spaces~\cite{bohm_dirac_1989, bohm_quantum_1998}, and so for the rest of this section when I say Hilbert space assume that I am talking about decomposing and factoring some kind of topological vector space with sufficiently nice properties.

\subsection{Hamiltonian-induced direct sum decomposition}

Observables can be used to partition the Hilbert space into subspaces. Each subspace is associated to an eigenvalue of the operator, and on that subspace that operator acts like a multiple of the identity.
The most familiar example is the Hamiltonian $H$. Assume for simplicity that the spectrum of the Hamiltonian $\sigma(H)$ is discrete, as it would be for most systems with finite extent. Then the Hilbert space for the system is decomposed on the spectrum of $H$ as
\begin{equation}\label{eq:HSinE}
\HS = \bigoplus_{E\in\sigma(H)} \HS_E.
\end{equation}
On the surface, this looks like (\ref{eq:HSdecomp}), but here instead of a decomposition into one-dimensional spaces, each energy eigenspace $\HS_E$ is realized by a finite-dimensional space $\mathbb{C}^{d(E)}$ with dimension equal to the degeneracy $d(E)$ of the energy $E$.

If the Hamiltonian under consideration is time-independent and describes a closed system, then the system has time-translation symmetry and the decomposition (\ref{eq:HSinE}) has another interpretation in terms of irreps. Time translation symmetry is an abelian symmetry, i.e.\ a system translated in time by intervals $t$ then $t'$ is the same as if translated by $t'$ then $t$.  Abelian symmetries have one-dimensional irreps and the unitary operators are just phases. Different irreps of time translation are distinguished by a scale-setting parameter (called the energy) that determines how fast the phase advances:
\begin{equation}
U(t')U(t)\HS_E = e^{-iEt'/\hbar}e^{-iEt'/\hbar}\HS_E = e^{-iE(t'+t)/\hbar}\HS_E = U(t'+t)\HS_E.
\end{equation}

When $\HS_E$ has more than one dimension, it is therefore a sum of multiple time-translation irrep spaces. This implies the existence of at least one other operator that commutes with $H$ and can be used to diagonalize the subspace $\HS_E$. Formally, one can always construct a single operator on the Hilbert space that commutes with $H$ and whose eigenvalues uniquely distinguish every vector in every degenerate space. Such an operator can be chosen as block diagonal, one block for each $\HS_E$ consisting of a $d(E)$-dimensional diagonal matrix with, for example, the numbers $1$ through $d(E)$ on the diagonal. Call this operator $D$. Then  $\HS$ can be reduced to a direct sum of one-dimensional subspaces on the joint spectrum of $H$ and $D$. In this sense, $H$ and $D$ are a complete set of commuting observables. However, the observable $D$ is defined by its construction as an operator on $\HS$ and has no fundamental origin as, for example, an observable defined by a measuring apparatus or the generator of a symmetry transformation. It is an example of a formal mathematical structure without physical interpretation. One goal when analyzing a system is to find operators that perform the same diagonalizing function as $D$, but have some other physical meaning, perhaps from kinematic symmetries.

For a particular Hamiltonian, the decomposition of $\HS$ into energy subspaces $\HS_E$ is fixed. However, consider a family of Hamiltonians from the same model that can be characterized by a variable parameter $g$. The subspaces $\HS_E$ may coalesce or split as $g$ is varied. The energy eigenspaces are called levels and I argue they function as interpretational units that are treated as `real' objects. One speaks of level `dynamics', for example, levels shifting, splitting, diverging, etc., but what is doing these actions is quite abstract: a subspace built from irreps in a model with a family of Hamiltonians.

\subsection{Observable-induced direct sum decomposition}

Any observable $\Lambda$ with a discrete spectrum $\lambda \in \sigma(\Lambda)$ can serve as the origin of a decomposition like (\ref{eq:HSinE}), not just the Hamiltonian:
\begin{equation}
\HS = \bigoplus_{\lambda \in \sigma(\Lambda)} \HS_{\lambda}.
\end{equation}
As before, the dimensionality of $\HS_\lambda$ is the degeneracy of the eigenvalue $\lambda$. To diagonalize this degeneracy, one can add additional observables and arrive at a set of commuting observables $\boldsymbol{\Lambda} = \{\Lambda_1, \ldots, \Lambda_k\}$ with discrete eigenvalues denoted $\boldsymbol{\lambda} = \{\lambda_1,\ldots,\lambda_k\} \in \sigma(\boldsymbol{\Lambda})$. Of course the Hamiltonian could be one of these $\Lambda_i$. A general decomposition can be written as
\begin{equation}\label{eq:HSbfL}
\HS = \bigoplus_{\boldsymbol{\lambda} \in \sigma(\boldsymbol{\Lambda})} \HS_{\boldsymbol{\lambda}}.
\end{equation}

Note that $\sigma(\boldsymbol{\Lambda})$ is the joint spectrum of all $k$ observables. For many models, the joint spectrum cannot be decomposed into the product of spectra. When it can (see \cite{harshman_symmetry_2016} for examples of separable three-body Hamiltonians), then the decomposition can be further reduced into independent sums
\begin{equation}\label{eq:silver}
\HS = \bigoplus_{\lambda_1 \in \sigma(\Lambda_1)} \cdots \bigoplus_{\lambda_k \in \sigma(\Lambda_k)} \HS_{\boldsymbol{\lambda}}.
\end{equation}
As discussed in \cite{harshman_symmetry_2016}, this kind of spectral separability distinguishes Hamiltonians with  `silver' and `gold' separability from `bronze' separability (this is a different notion of separability than topological separability; see more discussion in Sect.~\ref{sec:emerge} below).

\subsection{Symmetry-induced direct sum decompositions}

The two previous subsections used observables to decompose the Hilbert space into a direct sum of observable eigenspaces. Groups of symmetry transformations can be used for the same purpose, but now the subspaces are not necessarily eigenspaces. Instead, the subspaces are irreducible representation spaces, also known as modules, and carry a linear representation of the transformation group. A representation that is irreducible means that within the corresponding module, there are no invariant subspaces.

For specificity (and to avoid details that are important but technical) consider a group $G$ that has only finite-dimensional representations. Discrete finite groups, compact Lie groups and their combinations have this property. Denote the label for an irrep by a Greek letter in parentheses like $(\mu)$ and the corresponding irrep space with dimension $d(\mu)$ by $\mathcal{V}^{(\mu)}\sim\mathbb{C}^{d(\mu)}$. Then the Hilbert space can be broken into sectors labeled by $(\mu)$:
\begin{subequations}\label{eq:HSirrep}
\begin{equation}
\HS = \bigoplus_{(\mu)} \HS_{(\mu)}.
\end{equation}
Each sector $\HS_{(\mu)}$ is a tower of irreps spaces
\begin{equation}
\HS_{(\mu)} = \bigoplus_i \mathcal{V}^{(\mu)}_i.
\end{equation}
\end{subequations}

This may seem all a bit abstract, so here are a few examples. Perhaps the simplest and most familiar is the case of parity. Parity is realized by a the finite group of two elements $Z_2$. This abelian group has two irreps denoted $+$ for even states under parity and $-$ for odd states. So the Hilbert space can be divided into sectors of even and odd states $\HS = \HS_+ \oplus \HS_-$. Because the group $Z_2$ has only one-dimensional irreps, nothing more can be said except that $\HS_+$ and $\HS_-$ are built out of one-dimensional subspaces that are invariant under parity. If parity is a kinematic symmetry of a model, then all energy eigenstates are in one of those two sectors and the expectation value of parity is a dynamical invariant. If parity is a kinematic symmetry of a family of Hamiltonians, then varying the parameters of the family mixes states within a sector, but not across sectors, i.e.\ changing the control parameter does not change the parity of a state.

Another familiar example is rotational symmetry in three dimensions. The eigenvalue of the operator representing angular momentum squared $\hbar^2s(s+1)$ can be used to characterize irreps and irrep spaces. The spin $s$ come in two infinite series, non-negative integers and non-negative half-integers. A startling fact, called a superselection rule, is that a decomposition into rotation group irreps only consist of one of those two types, either integer or half-integer irreps. There can never be a superposition of states with integer and half-integer total angular momentum. That means in any quantum model the Hilbert space only has sectors of integer or half-integer irrep spaces.

A final example is the symmetry of particle permutations for $N$ identical particles, realized by the symmetric group $S_N$. Here the irrep decomposition provides a conceptual unit for analyzing the meaning of identical particles. Irreps of $S_N$ are labeled by positive integer partitions of $N$. For example, there are four partitions of $N=4$: $[4]$, $[31]$, $[22]\equiv [2^2]$, $[211]\equiv[21^2]$, and $[1111]\equiv[1^4]$. That means the Hilbert space for a model with four identical particles can be decomposed into five sectors:
\begin{equation}\label{eq:symdecomp}
\HS= \HS_{[4]} \oplus \HS_{[31]} \oplus \HS_{[2^2]} \oplus \HS_{[21^2]} \oplus \HS_{[1^4]}.
\end{equation}
The sectors have transformation properties dependent on their corresponding irrep. The irrep labeled $[4]$ is the trivial one-dimensional representation in which all permutations are represented by multiplication by $+1$. So this sector $\HS_{[4]}$ is appropriate for representing bosons, in fact it is the full scope of possibilities for four identical bosons. The irrep labeled $[1^4]$ is the one-dimensional totally-antisymmetric representations where odd permutations are represented by $-1$ and even permutations by $+1$. The corresponding sector $\HS_{[1^4]}$ is therefore where the fermions live. The three other sectors become necessary when considering particle with parastatistics (an exotic generalization between fermions and bosons) or more prosaically when considering particles with spin and spatial degrees of freedom, but fixing or tracing over either the spin or spatial degrees.

A sector like $\HS_{[4]}$ or $\HS_{[1^4]}$ is not a monolithic space; it is a tower of irreps space. Alternatively, it can be decomposed into energy subspaces, parity subspaces, etc. The mathematical structure of decomposition into irreps provides not just a technical tool for solving problems with identical particles, but also key conceptual unit for a minimal interpretation of what identical particles mean. This is discussed in more detail below in the specific discussion of the few-body model.

As noted above, the Hamiltonian itself is the generator for time translation symmetry, so the decomposition into energy subspaces (\ref{eq:HSinE}) is also an example of the irrep decomposition (\ref{eq:HSirrep}). Kinematic symmetries add more structure. Then each energy eigenspace is a sum of irrep spaces of the kinematic symmetry:
\begin{equation}\label{eq:HSEinmu}
\HS_E = \bigoplus_{(\mu),i} \mathcal{V}_{E,i}^{(\mu)}.
\end{equation}
A familiar example is the reduction of hydrogen energy levels into subspaces with fixed orbital angular momentum, doubled by the presence of spin. 
In principle, the direct sum in (\ref{eq:HSEinmu}) extends over all irreps of the kinematic group and there can be multiple copies of the same irrep as in the examples above. However, when there are multiple irreps with multiplicity, that usually signifies the presence of additional kinematic symmetries. In the hydrogen atom example, there is the additional $SO(4)$ symmetry that explain why subspaces with different orbital angular momentum have the same energy. 

Define the \emph{maximal kinematic symmetry group} $G_H$ of the Hamiltonian $H$ as the group such that every $\HS_E$ corresponds to a single irrep $(\mu)$ of $G_H$. When this group can be found, energy levels \emph{are} irreps of the maximal kinematic symmetry group, and this is a powerful tool for analysis of the model. It allows the physics of degeneracy to be handled in a systematic, algebraic fashion because the symmetry group provides all invariant observables necessary to diagonalize degeneracies. Other observables in the model can be characterized by their transformation properties under the group, simplifying calculations of expectation values, transition rates, and perturbation theory. Further, if $H$ is part of a family of Hamiltonians, then how $G_H$ changes with varying parameters determine how the energy levels (irreps) split and merge and how invariants are broken and reformed.

To close this section on decomposition, consider a \emph{dynamic} symmetry group $G$ with irrep labeled by $(\mu)$ representations. Since the symmetry group does not commute with the Hamiltonian, there is no necessary relationship between irreps of $G$ and $\HS_E$. However, one possibility is that each irrep of $G$ is decomposable in a sum of energy eigenspaces, i.e. the reversal of (\ref{eq:HSEinmu}):
\begin{equation}\label{eq:dynamic}
\mathcal{V}^{(\mu)}_i = \bigoplus_{E\in\sigma^{(\mu)}_i(H)} \HS_E,
\end{equation}
where $\sigma^{(\mu)}_i(H)$ is a purely symbolic shorthand for the spectrum of energies $E$ corresponding to the irrep $\mathcal{V}^{(\mu)}_i$ and depends on the Hamiltonian $H$ and the symmetry group $G$ in a model-specific way. In this case, $G$ is called a spectrum-generating group for the Hamiltonian $H$~\cite{wybourne_classical_1974}.

\section{Hilbert Space Arithmetic: Tensor Product Factorizations}

The other method of partitioning a model is through factoring the model Hilbert space into a tensor product structure. Any finite-dimensional Hilbert space $\HS\sim\mathbb{C}^d$ can be factorized into a $k$-fold tensor product
\begin{equation}
\HS = \HS_1 \otimes \HS_2 \otimes \cdots \HS_k \sim \mathbb{C}^{d_1} \otimes \mathbb{C}^{d_2} \otimes \cdots \otimes \mathbb{C}^{d_k}
\end{equation}
as long as $d_1 \times d_2 \times \cdots \times d_k$ is factorization of the positive integer $d$. Such a factorization is not unique; any unitary matrix in $U(d)$ that cannot be factorized into $U(d_1)\times U(d_2) \times \cdots \times U(d_k)$ defines another factorization with the same structure but different subspaces $\HS_i$. The situation with a separable but infinite-dimensional Hilbert space is even wilder. At least formally, Hilbert subspaces of any finite dimension can be factored off willy-nilly.

However, as with the generic decomposition (\ref{eq:HSdecomp}), such factorizations do not necessarily have any physical meaning, even within the limited ontology of a formal model. It is the allowed set of observables that distinguishes which factorizations that have functional, conceptual or interpretational value. In this section, several methods are presented in which a factorization of a model Hilbert space has an operational meaning in terms of observables. 

\subsection{Models of independent, distinguishable subsystems}

The most transparent case is when a model describes a system that is composed of denumerable, independent, distinguishable systems. This is the bottom-up approach to building a model. Consider each of these systems as a submodel with a Hilbert space $\HS_i$ and set of observables $\mathcal{A}_i$. The total Hilbert space is the tensor product of $k$ subspaces $\HS_i$
\begin{equation}
\HS = \bigotimes_{i=1}^k \HS_i.
\label{eq:HStensor}
\end{equation}
Pure states in $\HS$ can be classified as to whether they are separable or not separable with respect to this factorization. In this context, separable is used in the algebraic sense that a separable pure state $\kt{\psi}$ can be written at the tensor product of states $\kt{\psi_i}\in\HS_i$ as
\begin{equation}\label{eq:factor}
\kt{\psi} = \kt{\psi_1} \otimes \kt{\psi_2} \otimes \cdots \otimes \kt{\psi_k}.
\end{equation}
An entangled pure state is not separable and does not admit a factorization like (\ref{eq:factor}). To be clear, this is separable in a totally different sense than the topological notion of a separable Hilbert space admitting a decomposition into denumerable one-dimensional subspaces (\ref{eq:HSdecomp}), and it is also not equivalent to the separability of a differential equation discussed below. 

The set of observables denoted $\mathcal{A}_\oplus$ is constructed using the \emph{Kronecker sum} of the subsets of observables $\mathcal{A}_i$. To understand what that means, the Kronecker sum of two operators in different algebras $A\in \mathcal{A}_1$ and $B\in \mathcal{A}_2$  is
\begin{equation}
A \oplus B = A \otimes \mathbb{I}_2 + \mathbb{I}_1 \otimes B,
\label{eq:Kron}
\end{equation}
where $\mathbb{I}_i$ is the identity operator in $\HS_i$. This definition of the Kronecker sum can be generalized to more factors. (Unfortunately, note that the same symbol is typically used for the direct sum of vector spaces and the Kronecker sum of operators.)  For observables in $\mathcal{A}_\oplus$, expectation values of measurements factor into a sum of expectation values in each subsystem, and this holds for both separable and entangled pure states. I argue this can be taken as the definition of a model of independent systems. Note that $\mathcal{A}_\oplus$ is not an algebra of observables, even if each $\mathcal{A}_i$ were algebras. The set $\mathcal{A}_\oplus$ is closed under sums, but not under products.

Technically, a distinction is necessary between the sub-model observables $\mathcal{A}_i$ and the representation of those observables in $\mathcal{A}_\oplus$. The representation of $\mathcal{A}_i$ in $\mathcal{A}_\oplus$ is found by taking the Kronecker product of elements of $\mathcal{A}_i$ with the identity elements in all other subspaces $\HS_i$. In other words, the element $A \in \mathcal{A}_1$ is mapped into the element $A\oplus \mathbb{I}_2 \oplus \cdots \oplus \mathbb{I}_k$ in the set $\mathcal{A}_\oplus$. It is in this sense that we can say that the subsets of operators in the model commute with each other, another indication of their independence in this model.

Within each submodel is a Hamiltonian $H_i \in \mathcal{A}_i$. Considered as operators of the total space $\HS$, the sub-Hamiltonian $H_1$ is realized by $H_1 \otimes \mathbb{I}_2 \otimes \cdots \otimes \mathbb{I}_k$ and so on for all $k$ sub-Hamiltonians. The total Hamiltonian 
\begin{equation}\label{eq:KroneckerHam}
H = \bigoplus_{i=1}^k H_i
\end{equation}
describes a system composed of subsystems that are non-interacting. When $H$ is exponentiated $\exp(-iHt/\hbar)$ to generate time translations, it factors into a product of unitary operators on each subspace $\HS_i$
\begin{equation}\label{eq:KronTime}
U(t) = \exp(-i H_1 t/\hbar)\otimes\exp(-i H_2 t/\hbar) \otimes  \cdots \otimes \exp(-i H_k t/\hbar).
\end{equation}
Because time translation factors into operators that do not mix subspaces, entanglement is time-invariant. The only entanglement in the system is present in the initial state and remains unchanged. This definition can be lifted to described the tensor product structure in (\ref{eq:HStensor}): it is a dynamically invariant tensor product structure~\cite{harshman_galilean_2007}.

One physical interpretation of such a model is that the subsystems are on isolated patches of space. For example, in this case each Hilbert subspace $\HS_i$ might be a spin (i.e.\ an irrep of the rotation group). Even if these subsystems are internally identical, they can be distinguished by their location. In this sense, the factorization of the tensor product corresponds to the intuitive notion of \emph{locality}, the observables in $\mathcal{A}_\oplus$ are local observables, and the unitary operator is a local unitary. However, the subsystems could just as well be in the same location, but non-interacting and distinguishable. In this case, one could still use the term locality to refer to operators in $\mathcal{A}_\oplus$ that are local with respect to the tensor product of the structure. In fact, following Zanardi (see below) I have used this generalization of the term locality in numerous talks, including at the workshop that inspired this article, and received angry rebukes. A significant portion of the audience always seem to prefer that local retain its original meaning in terms of space (or space-time for relativistic systems). In response to the persistence head wind I have finally conceded and tentatively propose the terms \emph{specific} and \emph{specificity} to replace local and locality in this context.

\subsection{Models of interacting subsystems}

A standard assumption of quantum mechanics is that the model for an interacting system can be built from the models of the subsystems. The construction of the Hilbert space by tensor product is the same as before, and therewith follow the same notions of specificity and entanglement among subsystems. The difference is now that the set of observables is extended to include operators not in $\mathcal{A}_\oplus$. In the most extreme case, the total set of observables could be the algebra of observables is constructed as the tensor product of the subsets $\mathcal{A}_i$ as
\begin{equation}
\mathcal{A} = \bigotimes_{i=1}^k \mathcal{A}_i.
\end{equation}
More generally, the set $\mathcal{A}$ includes at least some observables that are not specific to the tensor product structure induced by the subsystems.

For an interacting system, the Hamiltonian must be non-specific, i.e.\ an operator that cannot be constructed by Kronecker products of sub-Hamiltonians like (\ref{eq:KroneckerHam}). As a result, time evolution no longer factors into a specific unitary operator like (\ref{eq:KronTime}). Entanglement of a state evolving in time is typically no longer a dynamical invariant with respect to the tensor product structure (\ref{eq:HStensor}). However, there may be other observables that are dynamical invariants. There may even be other tensor product structures besides the original construction that are dynamically invariant. The question becomes can one exploit these observables to find an alternate factorization and an alternate notion of specificity?

\subsection{Zanardi's theorem and virtual subsystems}\label{sec:zanardi}

For systems with finite-dimensional Hilbert spaces, one step towards answering this question is addressed by Zanardi's theorem~\cite{zanardi_virtual_2001,zanardi_quantum_2004,harshman_observables_2011}. It provides criteria for whether a partition of the observables leads to a tensor product structure and a notion of specificity.

A version of the theorem can be stated as follows:
\begin{quotation}
Given a state space $\Phi \subseteq \HS\sim\mathbb{C}^d$ and a collection of subalgebras $\{\mathcal{A}_1,\mathcal{A}_2,\ldots\}$ of the total algebra of observables $\mathcal{A}$ acting on $\HS$, the subalgebras induce a tensor product structure $\HS = \bigotimes_i \HS_i$ is they satisfy the following criteria
\begin{itemize}
\item \emph{Subsystem independence}: The subalgebras commute $[\mathcal{A}_i,\mathcal{A}_j]$ for all $i,j$.
\item \emph{Completeness}: The subalgebras generate the total algebra of observables $\mathcal{A} = \bigotimes_i \mathcal{A}_i$.
\item \emph{Specific (n\'ee local) accessibility}: Each subalgebra corresponds to a set of controllable observables.
\end{itemize}
\end{quotation}
In this statement, the first two requirements on the subalgebras of observables are mathematical in nature, and could be assessed from within the model as true or false for any particular partition of the observables. However, the third requirement is a physical criterion about empirical accessibility of measurement and control. There could be partitions of the observables that satisfy the first two, but are inadmissible based on a physical limitation of reality or some other constraint from outside the model.

An extension of Zanardi's theorem called the tailored observables theorem demonstrates the flexibility provided by the first two requirements in constructing subalgebras that factor the Hilbert space into `virtual' subsystems. For a finite-dimensional Hilbert space $\HS\sim\mathbb{C}^d$, one can construct  subalgebras of observables that induce a tensor product structure from a finite basis of operators such that any known pure state can have any entanglement that is possible for any prime factorization of $d$~\cite{harshman_observables_2011}. The proof relies on the unitary equivalent of Hilbert spaces with the same dimension and it is constructive in the sense that a procedure is given to construct the generators for the subalgebras in a finite number of steps (depending on the prime factorization of $d$). A consequence of this theorem is that for any pure state observables can be found that will detect as much or as little entanglement as is possible in a Hilbert space with dimension $d$. The only hitch is that entanglement is completely relative only when the control of the system is absolute, and therefore the third criterion of Zanardi's theorem is unrestrictive. For example, in a system of linear quantum optics (i.e.\ using only mirrors, phase shifters and beam splitters) any finite-dimensional unitary operator can be implemented~\cite{reck_experimental_1994}. Combined with Mach-Zener interferometers, this system has enough control to extract any observer-relevant entanglement from any pure state.

\subsection{Top-down approach to emergent tensor product structures}\label{sec:emerge}

As stated and proved, Zanardi's theorem and the tailored observables theorem hold for finite-dimensional Hilbert spaces and algebras of observables. It should have generalizations to separable Hilbert spaces and more general sets of observables, but I am unaware of any work in that direction.

Nonetheless, there are other cases of greater scope where partitions of observables provided by extra-model considerations lead to novel tensor product structures. Decoherence can select preferred tensor structures and thereby notions of subsystems, see \cite{jeknic-dugic_quantum_2013} and reference for a review. This approach to building the classical-quantum correspondence using decoherence and emergent tensor product structures has been referred to as the `top-down' approach~\cite{fortin_top-down_2016}.

Another category of top-down virtual subsystems is provided by yet another notion of separability: separation of variables and separation of integration constants. The Hamiltonian can be represented as the Schr\"odinger operator for models with an underlying $D$-dimensional space or configuration space. This kind of separability describes the existence of an orthogonal coordinate system that totally separates the Schr\"odinger equation into $D$ one-dimensional differential equations (or partially separates it into $d<D$ differential equations). There are different levels of separability depending on how the separated differential equations depend on the separation constants.

The `gold standard' is when each differential equation only depends on a single separation constant $\lambda_i$~\cite{harshman_symmetry_2016}. Then each differential operator $\Lambda_i$ defines a Zanardi-like subsystem and the whole Hilbert space can be decomposed as
\begin{equation}
\HS = \bigoplus_{\sigma(\lambda_1)} \cdots \bigoplus_{\sigma(\lambda_D)} \HS_{\lambda_1} \otimes \cdots \otimes \HS_{\lambda_D},
\end{equation}
where $\sigma(\lambda_i)$ is the spectrum of the differential operator $\Lambda_i$. This is a more robust separability than `silver' separability described in (\ref{eq:silver}), where only the spectrum was separable, not the Hilbert space. For `bronze' separability, the spectrum of each differential operator depends on the values of other separation constants and so spectral separability is lost. Only for gold separability does differential separability correspond to a tensor product structure and therewith an algebraic notion of separability.

A final method for identifying top-down tensor product structure is requiring that the tensor product structure be invariant with respect to a symmetry group of the model. A general theory of when this is possible has not been developed, but two examples from non-relativistic physics illustrate the idea.

Consider a model whose Hilbert space is an irrep space of the Galilean group and whose algebra of observables is the Galilean algebra extended by a central mass observable, i.e.\ a model of a single non-relativistic particle~\cite{levy-leblond_nonrelativistic_1967}. Elements of this irrep space are wave functions in three-dimensional space with an internal spin degree of freedom. The Hilbert space has a natural factorization
\begin{equation}
\HS = \HS_{\mathrm{space}} \otimes \HS_{\mathrm{spin}} \sim L^2(\mathbb{R}^3) \otimes \mathbb{C}^{2s+1}.
\end{equation}
Because it is an irrep of the Galilean group, no vector in this space is invariant under all transformations. However, this factorization of spatial and spin degree is invariant. The unitary operator that represents any Galilean transformation factors into a product of a specific unitary operator on each subspace~\cite{harshman_tensor_2007}. In contrast, irrep spaces of Poincar\'e transformations, corresponding to relativistic particles, do not have an invariant factorization between spatial and spin degrees of freedom, although they may under a subgroup of transformations~\cite{harshman_basis_2005}.

Another example is a model built from two Galilean irreps $\HS = \HS_1 \otimes \HS_2$ and a Hamiltonian with interactions that is not specific to that factorization, but for which the center-of-mass degrees of freedom still separate (in the sense of a differential equation). Then there is a alternate tensor product structure
\begin{equation}
\HS = \HS_{\mathrm{com}} \otimes \HS_{\mathrm{rel}}
\end{equation}
between the center-of-mass and relative degrees of freedom. This tensor product structure is symmetry-invariant with respect to Galilean transformation, and further, is it a dynamically-invariant tensor product structure. The unitary operators for a general Galilean transformation and for time-translation with the interacting Hamiltonian both factor into specific unitaries~\cite{harshman_galilean_2007}. As a result, entanglement between center-of-mass and relative degrees of freedom is invariant under transformation of coordinate systems and invariant in time. This kind of entanglement between center-of-mass and relative degree of freedom is not peculiar; it is present even in typical initial states of a scattering experiment where the particles are not originally entangled with respect to the interparticle tensor product structure $\HS = \HS_1 \otimes \HS_2$.

\section{Context: The Simplest Quantum Few-Body Problem}

The purpose of this section is to apply the techniques of Hilbert space decomposition and factorization to a specific quantum mechanical model describing a few interacting particles in one dimension. I consider distinguishable and indistinguishable particles without internal degrees of freedom like spin (or if they have spin, the spin is fixed and not dynamical). It is arguably the simplest quantum model that exhibits the full range of complexity of quantum dynamical systems. Understanding the model requires assessing the interplay of interaction, indistinguishability, identity, integrability, solvability and entanglement. A consistent, coherent synthesis of these issues provide a challenge for any interpretation of quantum mechanics in terms of units of reality. I claim that combining the bottom-up and top-down structural perspectives of model symmetries and Hilbert space arithmetic responds to this challenge with surprising depth.

\subsection{Single particle Hamiltonian}

A first step in building the interacting few-body model is describing the single-particle model. The Hamiltonian for a single particle in one dimension experiencing a static externally-generated potential:
\begin{equation}\label{eq:h1}
H^1 = - \frac{1}{2m}\frac{\partial^2}{\partial x^2} + V(x).
\end{equation}
Here is the first place the restriction to one-dimensional systems pays dividends. First, one dimensional systems are always integrable. An integrable system has as many algebraically-independent, globally-defined constants of the motion as the number of degrees of freedom~\footnote{The classical definition of integrability is usually formulated in term of operators generating flows on phase space. In a classical one-dimensional system, the constraint provided by this integral of motion reduces the two-dimensional phase space to a one-dimensional manifold, i.e. the trajectory of the particle. There is some ambiguity in the quantum case. See \cite{caux_remarks_2011} for a review of the difficulties.}. For a one-dimensional system, the Hamiltonian itself is the single conserved integral of motion necessary for integrability.
This does not mean that the system is necessarily solvable, in the sense that the eigenvalues and eigenstates of the Hamiltonian can be expressed in closed-form analytic expressions. However, for moderately well-behaved trapping potentials~\cite{loudon_one-dimensional_1959,harshman_infinite_2017}, Sturm-Liouville theory guarantees that the energy spectrum of (\ref{eq:h1}) is a denumerable tower of singly-degenerate states bounded from below. The wave function with lowest energy $\epsilon_0$ has no nodes, and each successive state with energy $\epsilon_n$ has $n$ nodes. 

Again assuming a reasonable trapping potential, the Hilbert space of the one-particle system $\HS^1$ is the space of Lebesgue-square-integrable functions $L^2(\mathbb{R})$ on the real line. This space carries an irrep of the one-dimensional Galilean group, although the trapping potential breaks that symmetry. Each energy eigenstate $\kt{\epsilon_n}$ spans a one-dimensional subspace $\HS^1_{\epsilon_n} \subset \HS^1$, leading to the decomposition of $\HS^1$ into a direct sum of energy eigenspaces (or equivalently, time translation symmetry irrep spaces):
\begin{equation}\label{eq:hs1}
\HS^1 = \bigoplus_{n=0}^\infty \HS^1_{\epsilon_n}.
\end{equation}

Note that in two and more dimensions, integrability is not guaranteed without more knowledge of the the potential and its symmetries. Although states can still be labeled by a spectrum of energies and a decomposition in energy subspaces like (\ref{eq:hs1}) is still possible, the subspaces are not necessarily one-dimensional and much less can be inferred about the properties of the wave functions.

\subsection{$N$ non-interacting, distinguishable identical particles}

The next step in constructing this minimal model is to combine $N$ non-interacting particles. 
The Hilbert space for the system is constructed by the tensor product of single particle Hilbert spaces
\begin{equation}\label{eq:HSN0}
\HS^N = \bigotimes_{i=1}^N \HS^1_i.
\end{equation}
One realization of this Hilbert space is as Lebesgue square-integrable functions on configuration space $L^2(\mathbb{R}^N)$. The Hamiltonian can be written as a sum of differential operators acting on functions of $N$ one-dimensional coordinates $x_i$:
\begin{equation}\label{eq:hN0}
H^N_0 = \sum_{i=1}^N \left( - \frac{1}{2m}\frac{\partial^2}{\partial x_i^2} + V(x_i) \right).
\end{equation}
The non-interacting dynamical model with Hamiltonian (\ref{eq:hN0}) is separable. Each degree of freedom $x_i$ is independent, and integrable; each single-particle Hamiltonian is an integral of the motion.  For distinguishable particles without interactions, the $N$ coordinates $x_i$ remain dynamically uncoupled. Though occupying the same physical space, the particles might as well be scattered throughout the galaxy as far as the dynamics are concerned. The Hamiltonian $H^N_0$ is specific to the tensor product structure (\ref{eq:HSN0}) and any entanglement among the particles is a dynamical invariant.

The kinematic symmetry group of $H^N_0$ includes the finite group of particle permutations isomorphic to $\mathrm{S}_N$. The transformation in this group are represented by operations on $\HS^N$, but they can also be realized as orthogonal transformations, i.e.\ reflections and rotations, on the configuration space $\mathbb{R}^N$~\cite{harshman_one-dimensional_2016}. This realization of particle permutations by geometrical point transformations connects back to the Platonic solids of the introduction. For example, the realization of $S_3$ in configuration space $\mathbb{R}^3$ is the point group of a triangle, the realization of $S_4$ in $\mathbb{R}^4$ is the point group of a tetrahedron, and so on for $N$-dimensional symmetries of $N$-simplices. If the single-particle system has parity symmetry (i.e.\ reflection symmetry about some point) then in addition a sequence of cubic-type symmetries appear~\cite{harshman_one-dimensional_2016-1}.

An energy eigenstate basis for the total system is formed by all tensor products of single-particle basis vectors like $\kt{\mathbf{n}} \equiv \kt{\epsilon_{n_1}}\otimes \kt{\epsilon_{n_2}} \otimes \cdots \otimes \kt{\epsilon_{n_N}}$. The energy $E_{\mathbf{n}}$ of a basis state is the sum of the single-particle energies. Most energies are no longer singly degenerate, but they are still denumerable and provide a decomposition of $\HS^N$ into a tower of energy eigenspaces
\begin{equation}\label{eq:HSN0irrep}
\HS^N = \bigoplus_{E_\mathbf{n}} \HS_{E_{\mathbf{n}}},
\end{equation}
where the direct sum is over all possible energies constructed as sums of $N$ single-particle energies $\epsilon_n$. Note that each space $\HS_{E_{\mathbf{n}}}$ has a complete basis that is unentangled with respect to the tensor product structure (\ref{eq:HSN0}).

The dimension of $\HS_{E_{\mathbf{n}}}$ is determined by the number of ways the set of single-particle energies that sum to $E_{\mathbf{n}}$ can be permuted. For example, for three particles the spaces $\HS_{E_{\mathbf{n}}}$ can have one, three or six dimensions. Irreps of $S_3$ either have one or two dimensions, and that signals the presence of additional kinematic symmetries beyond $S_3$~\cite{leyvraz_accidental_1997, fernandez_symmetry_2013}.

In fact, the decomposition (\ref{eq:HSN0irrep}) is a reduction into the irrep spaces of the kinematic symmetry group $T_t \wr S_N$, there $T_t$ is the time translation group of a single-particle system and $\wr$ is the wreath product~\cite{harshman_one-dimensional_2016-1}.  There could be an even larger kinematic symmetry group of the same form incorporating additional single-particle symmetries like parity.

Note that I have not made the claim that all the spaces $\HS_{E_{\mathbf{n}}}$ correspond to different energies. That depends one whether each energy can be uniquely associated to a set of $N$ single-particle energies.  If several energy sums coincide, then there must be an even larger kinematic symmetry group. I call this an emergent kinematic symmetry because it cannot be generated by single-particle symmetries and particle permutations. One example is when the trapping potential is a harmonic trap and then the maximal kinematic symmetry is $U(N)$ and can be realized as symmetry transformations on phase space~\cite{baker_degeneracy_1956, louck_group_1965}. For this system, the degeneracies of the energy eigenspaces grow like a factorial in the energy but can reduced into spaces like $\HS_{E_{\mathbf{n}}}$. Another example of an emergent kinematic symmetry is when the trapping potential is an infinite square well. Then there are `pythagorean degeneracies' that do not appear to have a description as a group of transformations realized on configuration space or phase space~\cite{shaw_degeneracy_1974}.

\subsection{$N$ non-interacting indistinguishable particles}

If the particles are indistinguishable fermions or bosons, then the model of $N$ non-interacting particles described above contains states and observables that cannot physically be realized or measured. There are several approaches to refining the model to account for indistinguishability, some more `bottom-up' and others more `top-down'.

One traditional bottom-up approach starts with the overcomplete Hilbert space $\HS^N$ and uses the symmetry group of particle exchanges to decompose it into sectors. Each sector is labeled by an irrep of the symmetric group $S_N$, and each sector is a tower of irrep spaces like (\ref{eq:HSirrep}). In the example given above with four particles (\ref{eq:symdecomp}), there were five kinds of irreps. More generally, there are as many irreps as there are positive integer partitions of $N$. The number $P(N)$ of inequivalent irreps  for a given $N$, or equivalently the number of partitions of $N$, is a combinatoric problem that does not have an algebraic expression. However, one of these is always the totally symmetric irrep $[N]$ which is one-dimensional and on which every particle permutation is represented by multiplication by $+1$. Another of these is the totally symmetric irrep $[1^N]$ which is also one-dimensional but now every odd particle permutation is represented by multiplication by $-1$ and the even permutations by $+1$.

Summarizing, the Hilbert space $\HS^N$ can always be decomposed as
\begin{equation}
\HS^N = \HS^N_{[N]} \oplus \HS^N_{[1^N]} \oplus \HS^N_{\mbox{everything else}}.
\end{equation}
The space $\HS^N_{[N]}$ contains all the allowed bosonic states and the space $\HS^N_{[1^N]}$ contains the allowed fermionic states. Further, because particle permutations are a kinematic symmetry of $H^N_0$, all of the energy eigenspaces $\HS_{E_{\mathbf{n}}}$ can be similarly decomposed. For one-dimensional systems, there is a single bosonic state in each $\HS_{E_{\mathbf{n}}}$, but there is only a fermionic state  in $\HS_{E_{\mathbf{n}}}$ when all the single-particle states are different.

The sectors  $\HS^N_{[N]}$ and $\HS^N_{[1^N]}$ do not inherit the tensor product structure (\ref{eq:hN0}) from the distinguishable particle construction. In fact for fermions, there are no pure states that are separable with respect to the tensor product structure (\ref{eq:hN0}). That sounds exciting on the surface, but because of indistinguishability there are also no observables that can detect the implied interparticle entanglement correlations of pure states~\cite{benatti_entanglement_2014}. One method to define entanglement in bosonic and fermionic systems is to find several complete commuting subalgebras of observables, a l\'a Zanardi, and use them to partition the symmetrized spaces $\HS^N_{[N]}$ and $\HS^N_{[1^N]}$.

A distinct, somewhat more `top-down' approach to indistinguishable particles builds the model on the symmetrized few-body configuration space~\cite{leinaas_theory_1977}. In the case of $N$ one-dimensional particles, one takes the quotient of the configuration space $\mathbb{R}^N$ with the symmetric group $\mathbb{R}^N/S_N$. This means that points in configuration space ${\bf x} = \{x_1,x_2,\ldots,x_N\}$ and ${\bf x}' = \{x'_1,x'_2,\ldots,x'_N\}$ that are equivalent up to a permutation of coordinates are identified as the same point. In this way, indistinguishability becomes a topological notion.  The topologically trivial manifold $\mathbb{R}^N$ is warped into a topologically non-trivial orbifold $\mathbb{R}^N/S_N$.
For one-dimensional and three-dimensional systems in Euclidean space, this approach leads to an equivalent formulation containing bosons and fermions. However, for two-dimensional systems, or other systems with topologically non-trivial one-particle spaces, this topological approach to identical particles can yield new physics, most famously anyons in two dimensions~\cite{wilczek_quantum_1982}.

\subsection{Contact Pairwise Interactions}

The final piece of the model adds contact (or zero-range) interactions between each pair $\langle i,j\rangle$:
\begin{equation}\label{eq:hN}
H^N = \sum_{i=1}^N \left( - \frac{1}{2m}\frac{\partial^2}{\partial x_i^2} + V(x_i) \right) + g \sum_{\langle i ,j \rangle} \delta(x_i - x_j).
\end{equation}
The contact interaction is given functional form by the one-dimensional delta-function weighted by the interaction parameter $g$. The contact interaction is appropriate for modeling physical scenarios where the range of the pairwise interaction is much shorter than any other length scale in the problem, e.g.\ the length scale determined by the trap and from the deBroglie wave lengths. The fact that the contact interaction is characterized by a single parameter makes it particularly amenable to analysis, as described below.

This model has been well-studied for over 50 years, going back at least as far as analyses for trapped  bosonic particles in the $g \rightarrow \infty$ limit by Girardeau~\cite{girardeau_relationship_1960}, free bosonic particles by Lieb and Liniger~\cite{lieb_exact_1963}, and free fermions by C.N. Yang~\cite{yang_exact_1967}. For reviews, including experimental implementations, see~\cite{cazalilla_one_2011,guan_fermi_2013}.

The Hamiltonian $H^N$ is no longer specific to the single-particle tensor product structure. Interactions break the separability and integrability of $H^N_0$ and engender correlations among the distinguishable or indistinguishable particles. However there are two important limiting cases.
\begin{itemize}
\item When $g \rightarrow \infty$, the interactions are called hard-core interactions. In one dimension, there is no way for particles to move past each other and so distinguishable particles would remain in a fixed order. In this limit, the system is no longer separable, but integrability reemerges for any trapping potential $V(x)$. As Girardeau showed, the wave functions can be expressed as algebraic combinations of the non-interacting wave functions restricted to specific orderings~\cite{girardeau_relationship_1960,harshman_identical_2017}. However, in this limit $H^N$ is not continuous and self-adjoint on all $L^2(\mathbb{R}^N)$, which creates some difficulties for level dynamics~\cite{sen_perturbation_1999}.
\item When the trapping potential is homogenous but finite in extent, i.e.\ the infinite square well, then the system is integrable for any value of $g$ using a method called the Bethe ansatz~\cite{batchelor_1d_2005, oelkers_bethe_2006}. There are $N$ integrals of motion that are symmetrized polynomials in the single-particle momenta.
\end{itemize}
Both of these limiting case can be understood as examples when the Yang-Baxter equation holds and there is diffractionless scattering~\cite{sutherland_beautiful_2004}.

One more special case is when the external trap is quadratic in position. The non-interacting system $H^N_0$ is equivalent to an isotropic harmonic oscillator in $N$ dimensions and, as mentioned above, has kinematic symmetry group $U(N)$. That system is maximally superintegrable, meaning there are $2N-1$ integrals of motion, and exactly solvable, meaning that the energy is an algebraic function of the quantum numbers and all excited states are products of the ground state with polynomials~\cite{post_families_2012}. At finite interaction strength, most of this additional analytical tractability is lost, but there is one extra integral of the motion corresponding to the separable center-of-mass degree of freedom~\cite{harshman_spectroscopy_2014}. This separability survives symmetrization of indistinguishable particles and therefore entanglement between center-of-mass and relative degrees of freedom remains a dynamical invariant.

For general traps and arbitrary $g$, the model (\ref{eq:hN}) is not integrable, nor is it solvable except numerically. Then two questions become: how far from integrability and deep into chaos and is it? How difficult is it to achieve convergent numerical solutions? The second question has been investigated exhaustively, by this author and many others, because of the relevance to current experiments~\cite{serwane_deterministic_2011, zurn_fermionization_2012, zurn_pairing_2013, wenz_few_2013}. For a partial list, see the references of \cite{harshman_one-dimensional_2016}. However, I would claim a productive metatheory of when particular approximation methods work well has not arrived.

One way to answer the first question about chaos is by comparing the spectrum to the Wigner-Dyson  distribution of eigenvalues a random matrix~\cite{gutzwiller_chaos_1990}. According to the Bohigas-Giannoni-Schmit conjecture, this is a universal feature of systems with quantum chaos. Work on closely related systems suggest that chaos is present in these systems~\cite{bohigas_characterization_1984}, but  the world currently waits for a more detailed analysis, especially one that situates the model in the hierarchy of chaos from ergodic, mixing, Kolmogorov, and Bernoulli~\cite{ullmo_bohigas-giannoni-schmit_2016, gomez_about_2017}.

In summary, depending on the trap shape and interaction strength, the model (\ref{eq:hN}) can manifest the full range of possible dynamic behaviors, from (super)integrability to (conjectured) hard chaos. For integrable cases, there are observables privileged by dynamical conservation laws that fully characterize the system, i.e.\ a complete set of commuting observables. The specific nature of the integrals of motion depend on the trap and interaction strength. For the non-interacting case, the integrals of motion are single-particle observables. In integrable interacting cases like the hard-core limit or Bethe-ansatz solvable cases, the conserved quantities are collective observables built from symmetrized polynomials of single-particle observables. In contrast, for chaotic cases, the unique conserved observable is the energy itself, and the spectrum looks matches a relevant form of randomness. A goal of this avenue of research is to see if the approach to chaos can be understood as the dissolution of structures based on irreps of symmetry group.

\section{Conclusion: A Few Comments on Symmetry, Structures and Solvability}

The previous section introduces \emph{in situ} certain technical terms of art like integrability and solvability. The reader should not be misled by my breezy usage of the terms to infer that there is universal acceptance among the community of mathematical physicists on how these terms should be applied. For example, Liouvillian integrability is naturally defined in classical dynamics on phase spaces, but there is debate on how is works for quantum systems with a mixture of continuous and discrete degrees of freedom~\cite{caux_remarks_2011}. And there is not consensus on the relationship between Bethe-ansatz integrability relevant to the model above and Liouvillian integrability and I know only a few case where they coexist, e.g.\ \cite{harshman_integrable_2017}. Even more egregious is the dissent and dissemblance around the term solvable. Some physicists  throw around the term `exact solution' when they have actually found a somewhere-convergent, asymptotically-approximate numerical solution to the limiting case of a mathematically ill-conditioned dynamical model.

I argue that solvability should be considered as a property of a dynamical model, and should be considered as a continuum. At the most solvable extreme are models where one can push analysis deep into the realm of pure algebra. Examples includes superintegrable systems and exactly solvable systems (which in fact are conjectured to be the same thing). Moving down the spectrum we have systems whose solutions formulated in exact analytic relations, but those relations may require (for example) solving transcendental equations for the spectrum or other model parameters. 

My argument is that this same continuum from more solvable to less solvable coincides roughly with two other features: (1) the amount of symmetry in the system, as measured by the size or complexity of the symmetry group; and (2) the tractability of Hilbert space arithmetic, meaning the variety of inequivalent ways the system can be decomposed or factorized into subspaces. The structural unit that unifies these two features is the irreducible representation.

Irreps appear in many guises: as invariant subspaces in direct sums and tensor products, as the building blocks of towers for describing identical particles that generate unusable entanglement and frustrate algebraic separability, as the concept of energy levels that vary across families of Hamiltonians in the same model, and more. No matter the underlying ontological commitment of an interpretation, any formulation of quantum mechanics must account for the prevalence and utility of these structures. Unlike the Platonic solids, these are not metaphors. They are mathematical building blocks of quantum mechanics and will remain so even when new or reformed ontologies emerge.

The most egregious oversight of this article is that I have not discussed how symmetry groups partition the set of observables into irreps. This is more technically challenging that the Hilbert space arithmetic I have presented, but I think the potential rewards are a deeper understanding of the connections among the observables, separability (in all three senses) and integrability.

At this stage, the investigation is still incomplete, but I argue that one notion emerges: the importance of solvability. Solvability is a concept lying in that awkward place of being a technical term with multiple overlapping and conflicting definitions in different contexts. One unifying themes across these contexts is that solvable systems play a central role in the interpretation of physical phenomena. Solvable models in mechanics, like coupled harmonic oscillators and hydrogenic atoms, are not just touchstones for mathematical analysis. They are ubiquitous as direct and approximate models in mature, and they provide the cognitive framework for understanding other physical systems. Is it a coincidence that solvable models are so useful? Is it just attention bias, i.e. we pay more attention to things we understand more thoroughly? Or is there something more deeply `real' about solvable systems, either in an epistemic or ontic sense?

\section*{Acknowledgments}

I would like to thank Olympia Lombardi and the other organizers of the workshop for assembling such a stimulating group of physicists and philosophers.

\end{document}